\documentclass[12pt,a4paper]{article}
\usepackage[T1]{fontenc}
\usepackage{indentfirst}
\usepackage[english]{babel}
\usepackage{amsfonts,latexsym}
\usepackage{amsthm}
\usepackage{euscript}
\usepackage{amsmath}
\usepackage{color}
\usepackage{graphicx}
\usepackage{caption}
\usepackage{subcaption}
\numberwithin{equation}{section}
\usepackage{ucs}
\usepackage[utf8x]{inputenc}
\usepackage{amsfonts}
\usepackage{graphicx}

\newfont{\bcb}{msbm10 scaled 1200}
\newfont{\bcc}{msbm10}

\title{New Proposals of a Stress Measure in a Capital and its Robust Estimator}
\author{Tadeusz Klecha$^1$, Daniel Kosiorowski$^1$, Dominik Mielczarek$^2$,
\\ Jerzy P. Rydlewski$^2$}
\begin{document}
\maketitle
\begin{center} 
$^1$\textit{Cracow University of Economics, Faculty of Management, Poland}
\\ $^2$\textit{AGH University of Science and Technology, Faculty of Applied Mathematics, Krakow, Poland;} 

\end{center}

\textbf{Abstract}
In this paper a novel approach for a measurement of stresses in a capital, which induce the capital flows between economic systems, is proposed. The proposals appeal to an apparatus offered by the statistical theory of shape. We propose a stress functional basing on a concept of mean shape determined by representative particles of a capital carrier. We also propose methods of describing changes in an amount and a structure of stresses in a capital appealing, among others, to a Bookstein's pair of thin plain spline deformation, and a measure of a shape variability. We apply our approach to an indirect verification of the hypothesis according to which a capital flow between economic systems is related to an activity of an inner force related to stresses in a capital. We indicate, that the stresses create a phenomenon analogous to the heat, which may be interpreted in terms of a positive economic external effect, which attracts a capital from environment of a system to the system. For empirical studies we propose robust approach to estimate the stress functional basing on the data depth concept. In the empirical research we use data on five branch stock indexes from Warsaw Stock Exchange. The studied period involves the financial crisis of 2007.
\\ \textbf{keywords:}
a stress functional, an average shape, shape variability, a deformation of objects, robust estimator of an average shape
\\ \textbf{JEL Classification: C13, C43, C63 }
\section{Introduction}

In the economics during last 30 years one could observe an intensive development of economic disciplines basing on advances in a theory of stochastic processes and mathematical statistics. Due to a development of various new models in an empirical finance, many academic economists believed, that we are able to better and better describe and forecast new financial phenomena like conditional heteroscedascity or regime switching in exchange rates (see Tsay, 2010 \cite{Tsay}). Due to advances in financial mathematics many analysts believed, that we are able to better and better value options for buying or selling stocks, raw materials, real estates.\\

The financial crisis of 2007 clearly showed limitations of a solely "mechanical", non-theoretical  approach to forecasting of economic systems behaviours. From other point of view, a significant part of econometric, data analytic and statistical community indicated a lack of precision as to empirical consequences of theories, a lack of operational versions of theories, an usage of "a magic" - undefined or unmeasurable terms like a "technological development", "productivity of a production factor" (see Romer, 2012 \cite{Romer}). Practitioners charge theoreticians with a triviality of conclusions like "workers move from places of low wages to places of high wages", and statisticians with taking unrealistic assumptions of stationarity, ergodicity, or commonly using asymptotic arguments. In this context one may notice an appearance of various new conceptual approaches for explaining past and predicting future economic crashes (see for example Lee and Nobi, 2017 \cite{Networks_crisis}).\\
There are also a more classical approaches to explanation of the upcoming crisis.
For example, Youssefmir et al. (1998) \cite{Youssefmir} constructed a simplified model of bubble formation and bursting. They showed that when speculative trends dominate over fundamental pricing, or expectations are trend chasing, it leads to growth of asset prices away from their fundamental value. Then bubbles form, what makes the financial system susceptible to any shock, leading eventually to a financial crash. The authors justified the mechanism of outcoming financial crisis, but do not bind it directly with the stresses in the capital.
\\ From other point of view, several recent data-analytic studies justify our efforts to incorporate by default nonlinear theory into financial crushes modelling. The authors showed (e.g. see Anderson
and Vahid (1998) \cite{anderson}, Kiani (2011) \cite{Kiani}, and Kiani and Bidarkota (2004) \cite{KianiBidarkota}) that most macroeconomic time series are nonlinear, and  hence linear models should not be used for forecasting fluctuations in economic activity.\\

In their paper Hakkio and Keeton (2009) \cite{KCFSI} qualitatively explained, how they understand financial stress, but no precise definition was set up. Nevertheless, they pointed the key features of financial stress: increased uncertainty about fundamental value of assets, increased uncertainty about behavior of other investors, increased asymmetry of information between lenders and borrowers or buyers and sellers of financial assets, and decreased willingness to hold risky or illiquid assets. They introduced a new measure of financial stress — the Kansas City Financial Stress Index (KCFSI). The measure is based on 11 financial market variables, each of which captures one or more considered key features of stress. The variables are designed for available in the United States financial market characteristics, however some of them, i.e., cross-section dispersion of bank stock returns, can be considered for any institutionally mature financial market. 
Coefficients of these 11 variables in the index were calculated and scaled. These
coefficients are chosen so that the index explains the maximum possible
amount of the total variation in the 11 variables.
The KCSFI allows governing bodies to gain some insight into the level of financial stress by comparing the current value of the KCFSI to its value in the past.\\
Federal Reserve Bank of St. Louis Financial Stress Index (STLFSI) is using the first principal
component of 18 weekly data series comprising financial variables:
seven interest rate series, six yield spreads and five other indicators related to global financial markets. The STLFSI is designed to measure global investment climate, and the strain imposed on international financial markets. 
\\ Dua and Tuteja (2016) \cite{Dua_Tuteja} investigated the impact of STLFSI on the correlation across financial markets and made some conclusions for strategies in the international equity and currency markets. 
\\ Our paper fits into a general research approach, which binds economic crashes with certain kind of stresses existing on markets.
We conducted empirical research on Polish financial market, and thus Warsaw Stock Exchange Index (WIG) and its subindices were considered. In our approach, for stresses evaluation we directly appeal to an origin of the stress concept, i.e., to theory of elasticity (see Landau (1953) \cite{Landau}) and propose a qualitative framework for a capital flows description basing on the rationale thermodynamics proposed in Truesdell (1969) \cite{Truesdell}. We show, that the approach may be useful in a prediction of an approaching crisis.\\
We apply the approach to a verification of a hypothesis stating that \emph{capital flows between economic systems are closely related to an activity of inner stresses between particles of the capital carrier}. In our proposals we translate the approach into a language of a statistical theory of shape (STS). For the verification of the posted research hypothesis we robustify tools of the STS using procedures offered by modern discipline of robust multivariate statistics called the data depth concept (see Zuo and Serfling, 2000 \cite{Zuo_Serfling}). We believe, that the framework enables us for a better understanding the root cause of a crisis.  

From a strictly statistical point of view, one should notice, that economic datasets are generated by time varying models with various complex ways of varying (Anderson and Vahid, 1998 \cite{anderson}, or Rydlewski and Snarska, 2014 \cite{CHARME}), which are often too complex for reasonable stochastic description and modeling and estimation (for example due to the course of dimensionality). Moreover, observed datasets are noisy and often consist of outliers of various type (see Galeano et al., 2006 \cite{Galeano_et_all}, Kosiorowski, 2012a \cite{Kosiorowski_compstat_12}). 
Thereby, we would like to point out an importance of the preliminary analytical step consisting of proper selecting variables, and transforming or cleaning of the data. On the other hand, we believe that better theoretical understanding of the economic systems would allow us for a simplification of the statistical issues under study leading to a better specification of a model used for prediction, monitoring, and managing the economic systems.\\

The rest of the paper is organized as follows. In Section 2 a general conceptual setting is sketched, and in Section 3 a proposal of a stress functional in a capital is presented. Section 4 consists of our proposals of robust approach to estimation of a stress functional, and a strain force causing an outflow of a capital from a system. Section 5 presents selected results of empirical studies conducted within the proposed framework. Paper ends with conclusions and references.   

\section{General setting }
A notion of a capital belongs to basic terms of theoretical economics. It is an immanent part of various growth, business cycle, inflation or asset pricing models (Romer, 2012 \cite{Romer}). It should be stressed, however, that in a great part of economic models, one assumes that an amount of capital in a system relates to amounts of machines, raw materials, stocks, certified skills, diplomas of workers, etc., expressed in a certain currency (see Romer, 2012 \cite{Romer}, G\"ortz and Tsoukalas, 2017 \cite{Business}).

It is obvious however, that one can imagine, e.g., two software companies equipped with machines of similar values measured in dollars, with staffs possessing similar certificates - but with different abilities to expand, and conquer new markets.\\
A size of resources, expressed in currency, stored in an economic system certainly relate to its behavior and interactions with an environment, but certainly does not exhaust possibilities of understanding of a capital and its role even in the "classical" theory of economics (see Romer, 2012 \cite{Romer}, G\"ortz and Tsoukalas, 2017 \cite{Business}).\\ 
In 2005 Klecha and Kosiorowski formulated a concept, according to which, \emph{an amount of capital, stored in a certain economic system, is described by the ability of this system to perturb a certain space of economic values (a scalar field of economic values), a flow of the capital is an effect of an activity of forces related to both internal stresses in a substance of the capital carrier and external forces associated with activities of other systems (environment)}. Due to appearance of stresses in the capital carrier, a phenomenon analogous to physical heat production is observed. The phenomenon may be  interpreted in terms of a positive economic external effect influencing the environment, which among others attracts a capital from an environment of a system to a system. (see Fleurbaey and Maniquet, 2011 \cite{welfare}). Authors pointed out that one can neither observe directly the capital nor one can directly observe perturbations in the space of values. Nevertheless, we can monitor perturbations of interplay between demand and supply, that is,  perturbations in a space of prices. Within the concept, the capital should be treated as an energy stored in the economic system. A measure of an amount of the capital in the system should be closely related to an ability of the system to work, and with an amount of its influence on the space of economic values.
\\ The concept originates from Klecha (1996a \cite{Klecha1},   1996b \cite{Klecha2}, 2002 \cite{Klecha2002}). In Kosiorowski (2006) \cite{KosiorowskiPhD} the concept has been translated into economic language, and statistical apparatus for an indirect verification of the concept has been proposed (see Kosiorowski, 2007 \cite{KosiorowskiLisbon}). In this paper we briefly sketch the results obtained by Kosiorowski (2006) \cite{KosiorowskiPhD} and present an application of an improved robustified statistical apparatus appealing, among others, to the robust functional data analysis, i.e., data depth concept for functional data (for other applications of the data depth concept see Kosiorowski et al., 2017a \cite{StatPapers} and 2017b \cite{ComputStat}).
In order to verify the general concept, in Kosiorowski (2006) \cite{KosiorowskiPhD} the following assumptions were chosen.
\\ 1. Capital flows from a place of a smaller perturbation in the space of economic values to a place of a bigger perturbation in that space. 
\\ 2. The flow of the capital perturbs the space of values.
\\ 3. The flow of the capital can be caused by external and/or internal forces. 
\\ 4. An activity of inner forces in the capital is related to stresses between its particles. 
\\ 5. The stresses manifest in local changes of relative positions of representative particles of the capital carrier.
\\
In this paper we focus our attention on measurement of the inner force in the capital, which is related to the stresses between selected particles of its carrier. The force causes a capital flow between economic systems. We assume, that the force is determined by a value of the stress functional of a capital. As a starting point for the stress functional definition we take general constitutive equations (see Truesdell, 1969 \cite{Truesdell}). It means, that we lay down the following postulates on the stress functional.
\begin{itemize}
\item[1.] \emph{Principle of determinism for the stress}: the stress in a system is determined by the history of the motion of the system.
\item[2.] \emph{Principle of a local action}: in determining the stress at a given particle $x$, the motion outside a certain neighborhood of $x$ may be disregarded.
\item[3.] \emph{Principle of material frame indifference}: two observers considering the same movement in the system are noticing the same stress functional value.
\end{itemize}
Due to properties of the proposed stress functional and statistical properties of its estimator, i.e.,  an affine equivariance of the estimator, the above postulates are automatically fulfilled.

 \section{Proposals of the stress functional}
According to D. G. Kendall, a shape of an object is all the geometrical information that remains invariant, when location, scale and rotational effects are filtered out from the object (Dryden and Mardia, 2016 \cite{DrydenMardia}). Within statistical analysis of shape, objects of a considered population are studied on base of $m$, $k$-dimensional \emph{landmarks} (indicators, markers), which are points placed on objects, corresponding with certain essential mathematical or content-related properties of these objects. We represent an object by means of a configuration matrix, consisted of coordinates of the landmarks. In order to remove a location, we multiply the configuration matrix by a Helmert submatrix. In order to remove a scale, we divide the resulted matrix by its size, e.g., a centroid size, and then we obtain \emph{a preshape} of the object. Two preshapes represent the same shape if they differ by a rotation (for details see Dryden and Mardia, 2016 \cite{DrydenMardia}, and Goodal and Mardia, 1993 \cite{Goodal}).
\\ Let us consider two $n\times p$ matrices $X_1$ and $X_2$ consisting of coordinates of  $n$ landmarks placed on $p$ dimensional objects $X_1$ and $X_2$. Consider a following optimization problem
\begin{equation}
\min_{m,A}||X_2-X_1(A+1_pm^T)||_F,
\end{equation}
where $X_1$ and $X_2$ are both $n\times p$ matrices of corresponding points, $A$ is an orthonormal $p\times p$ matrix, $1_p$ is a column vector of ones, $m$ is a $p$-vector of location coordinates, and $||X||_F^2=trace(X^TX)$ is the squared Fr\"obenius matrix norm.
\\ Let $\overline{x}_1$ and $\overline{x}_2$ be the column mean vectors of the matrices, and \~X$_1$ and \~X$_2$ be the versions of these matrices with the means subtracted. Consider the SVD (singular values decomposition) \~X$_1^T$\~X$_2=UDV^T$, where both $U$ and $V$ are orthogonal matrices. It can be shown that the solution to (3.1) is given by $\hat A=UV^T$, $\hat m=\overline{x}_2-\hat A\overline{x}_1$ The minimal distance is referred to as the Procrustes distance. From the form of the solution, we can center each matrix at its column centroid, and then ignore location completely.
\\ The Procrustes distance with scaling, solves a slightly more general problem
\begin{equation}
\min_{\beta, \textbf{A}}||X_2-\beta X_1\textbf{A}||_F
\end{equation}
where $\beta>0$ is a positive scalar. The solution for $\hat{\textbf{A}}$ is as before, with $\hat{\beta}=\frac{trace D}{||X_1||^2_F}$.
\\ Strictly related to Procrustes distance is the Procrustes average of a collection of L shapes (configuration matrices $X_1,...,X_L$), which solves the problem
\begin{equation}
\min_{\{\textbf{A}_l\}_{l=1}^L,M}\sum_{l=1}^L||X_l\textbf{A}_l-M||_F^2.
\end{equation}
That is, find the shape $M$ closest with respect to average squared Procrustes distance to all the shapes. The problem my be solved numerically using free R package \emph{shapes} ( Dryden and Mardia, 2016 \cite{DrydenMardia}). It is worth noticing, that we can only expect a solution up to a rotation.
\\ In order to obtain a general measure of shape variability, it is convenient to use the root of the average square of the distance between each configuration and the Procrustes average:
\begin{equation}
SVAR=\sqrt{\frac1L\sum_{l=1}^L||X_l-M||_F^2}.
\end{equation}
In order to express a relation between stresses in the capital and capital flows Kosiorowski (2006) \cite{KosiorowskiPhD} analyzed stocks belonging to a sector stock index considered with respect to the price and the volume. We treated the stocks as representative particles (markers, landmarks) of the capital. We assumed that the average shape of the sector stocks index corresponds to the stress functional and a measure of the shape variability corresponds to a value of a force inducing a capital flow between economic systems. Within the research framework, we considered the following general systems of Polish economy: banking, construction, media, food industry, fuels (oil\&gas) industry, telecommunication industry, information technology industry (IT) and chemical industry in the period 29.12.2005 – 30.09.2011 basing on behavior of the corresponding branch indices in this period, i.e., WIG-banking, WIG-construction, WIG-media, WIG-food, WIG-oil\&gas, WIG-telecom, WIG-IT and WIG-chemicals. 

Note, that introducing the stress functional basing on mean shape of a system, we obtain a global measure of the stresses. However, in order to indicate a direction of an outflow/inflow of a capital induced by the stresses, we need a tool possessing a “more local” nature. For this reason, we propose to use a transformation called a deformation in a machine learning terminology. We propose to study properties of a deformation of two average shapes estimated for the two consecutive periods.    
\\ Let us briefly recall, that deformation in a continuum mechanics is the transformation of a body from a reference configuration to a current configuration. A configuration is a set containing the positions of all particles of the body. Deformation is the change in the metric properties of a continuous body. Deformation is usually caused by external loads, body forces or temperature changes within the body. Deformations which are (are not) recovered after the stress field has been removed are called elastic (plastic) deformations. A strain is a description of deformation in terms of relative displacement of particles in the body. Strains measure, how much a given deformation differs locally from a rigid-body deformation. One can describe the strain as a normalized measure of deformation representing the displacement between particles in the body relative to a reference length. In our setting, we propose to take the $SVAR$ measure of shape variability (see equation (3.4)) as a quantity being in a close relation to the strain. 
\\ In order to describe the difference in shapes of objects and to visualize the strain, we compute a transformation of the space in which the first object lies (stress functional calculated for a first period) into the space of the second object (stress functional calculated for the second period). The transformation gives us information about both the local and global shape differences. We use a Bookstein (1989) \cite{PTPS} thin-plane splines deformation, which decomposes a deformation into a global affine transformation and a set of local deformations, which highlight changes at progressively smaller scales. This decomposition is in close analogy with a Fourier series decomposition, with the constant term in Fourier series being the global parameter and the coefficients of the trigonometric terms being local parameters at successively smaller scales.
\\ Consider two $k\times m$ configuration matrices consisting coordinates of k-points in $\mathbb{R}^m$, $\mathbf{T}=(t_1,...,t_k)^T$, $\mathbf{Y}=(y_1,...,y_k)^T$. Suppose we wish to deform $\mathbf{T}$ into $\mathbf{Y}$. A deformation is a mapping from $\mathbb{R}^m$ to $\mathbb{R}^m$
defined by the transformation
\begin{equation}
\Phi : \mathbb{R}^m \ni t \longrightarrow \left(\Phi_1(t),\Phi_2(t),...,\Phi_m(t)\right)^T\in\mathbb{R}^m. 
\end{equation}
We shall concentrate on the most important for our purposes $m=2$ dimensional case, with deformation given by the bivariate function $y=\Phi(t)=\left(\Phi_1(t),\Phi_2(t)\right)^T$ which is: continuous, smooth, bijective, not prone to gross distortions, equivariant under transformations of relative location, scale and rotation of the objects, and is an interpolant, i.e., $y_j=\Phi(t_j).$
\\ Bookstein (1989) has proposed an approach for planar deformations. His pair of thin-plate splines (PTPS) is given by a bivariate function 
\begin{equation}
\Phi(\textbf{t})= \left(\Phi_1(\textbf{t}),\Phi_2(\textbf{t})\right)^T=\textbf{c}+\textbf{At}+\textbf{W}^Ts(\textbf{t}),
\end{equation}
where $\textbf{t}\in\mathbb{R}^2$,
$s(\textbf{t})=\left[\delta(t-t_1),\delta(t-t_k)\right]^T$ and 
\begin{displaymath}
\delta(h)= \left\{ \begin{array}{ll}
||h||^2\log\left(||h||\right), & ||h||>0\\
0, & ||h||=0.
\end{array} \right.
\end{displaymath}
The PTPS deformation has $2k+6$ parameters: scalar $\textbf{c}(2\times1)$ has $2$ parameters; affine transformation matrix $\textbf{A}(2\times2)$ has $4$ parameters, and “local” transformation matrix $\textbf{W}(k\times 2)$ has $2k$ parameters. There are $2k$ interpolation constraints: 
$\left[y_j\right]_r=\Phi_r\left[t_j\right]$, $r=1,2$; $j=1,...,k$, and 6 more constraints for a "bending energy": $\textbf{1}_k^T\textbf{W}=\textbf{0}$ and $\textbf{T}^T\textbf{W}=\textbf{0}$.
\\ The PTPS minimizes the amount of bending in transforming between two configurations (constraints on “bending energy” can also be treated as a roughness penalty, see Hastie et al., 2009, for details of theory of splines with constraints). It can be proved that PTPS transformation minimizes the total bending energy of all possible interpolating functions mapping from the configuration $\textbf{T}$ to the configuration  $\textbf{Y}$, where the total energy is given by:
\begin{equation}
J(\Phi)=\sum_{j=1}^2\int\int_{\mathbb{R}^2}\left[
\left(\frac{\partial^2\Phi_j}{\partial x^2}\right)^2+2\left(\frac{\partial^2\Phi_j}{\partial x\partial y}\right)^2+
\left(\frac{\partial^2\Phi_j}{\partial y^2}\right)^2\right]dxdy.
\end{equation}
Unfortunately, the PTPS is very sensitive to the input and output objects. One cannot directly use this tool to economical shapes estimated from the real data. For our purposes we need robust estimators of economic shapes (stress functional values), and the measure of a variability of shape (a quantity closely tied with the strain, $SVAR$).
\section{Proposal of a framework for robust analysis of capital flows between economic systems}
Unfortunately, statistical models commonly used within statistical analysis of shape, impose very high restrictions concerning the assumptions for the examined phenomenon. We mean a multivariate normality or even isotropy assumption for probability distributions generating configurations. The commonly used estimators are not robust (e.g. Procrustes average is in fact the least squares estimator, and it is a well known fact, that least squares estimators are not robust). The problem of finding a robust estimator of an average shape is not trivial, because it may happen that commonly used operations like trimming, Winsorising of configuration matrices “by coordinates” lead to an object object which does not represent any shape. Therefore, we have to define an estimator directly in a non-Euclidean shape space (see Goodal and Mardia, 1993), or act on configurations matrices in an appropriate way, which takes into account their multidimensional features. That facts motivate us to propose a robustification (Huber and Ronchetti, 2009 \cite{HuberRonchetti}, Maronna et al., 2006 \cite{Marona}) of Procrustes analysis appealing to an approach of modern robust multivariate statistics called \emph{data depth concept} (see Zuo and Serfling, 2000 \cite{Zuo_Serfling}). 
The concept originates from Tukey (1975) \cite{Tukey1975}, who introduced the notion of data depth and depth contours in order to visualize multivariate data and generalize one dimensional statistical techniques basing on ranks and quantiles to a multidimensional case.\\
In order to choose representative particles of a capital (understood according to our definition) we focused our attention on companies possessing "median trajectory" in the considered period with respect to a ratio between standardized price and standardized volume. For finding the median trajectories we used modified band depth of L\'opez-Pintado and Romo (MBD, see L\'opez-Pintado and Romo, 2009 \cite{LopezRomo}).
\vskip1mm
MBD of function $x$ with respect to functional sample $X_N=\{x_1,...,x_n\}$ is:
\begin{equation}
MBD(x|X^N)=\frac{2}{n(n-1)}\sum_{1\leq i_1< i_2\leq n} \frac{\lambda(A(x;x_{i_1},x_{i_2}))}{\lambda([0,T])},
\end{equation}
where $A(x;x_{i_1},x_{i_2})=\{t\in[0,T]:  \min_{r=i_1,i_2} x_{r}\leq x(t)\leq \max_{r=i_1,i_2} x_{r} \},$ and $\lambda$ denotes a Lebesgue measure. 
Evidently, MBD is a functional depth, which takes into account a proportion of "time", when a function $x$ is in the band made with any two functions from the considered functional sample $X_N$.
\vskip1mm
Functional depth function, e.g. MBD,  takes values in the interval $\lbrack 0,1 \rbrack$, compensates for lack of a linear order in function space, and provides a center-outward ordering of the functions with higher values representing greater “centrality” (a value close to one). Maximum depth functions define a notion of “center” and a notion of “functional median”. The set of functions for which depth function takes value not smaller than $\alpha$ is called $\alpha-$central region, 
and is treated as an analogue of one-dimensional quantile.
The nested regions can be constructed, that is, regions $\{ x: MBD(x|X^N)\geq \alpha\}$. The median with respect to the considered functional depth is the most central observation. We can also define a sample median as 
\begin{equation} MED_{MBD}(X^N)={\mathop{\arg \max }}_{i=1,...,N} MBD(x_i|X^N).
\end{equation}
Conventionally, if more than one function is achieving the depth maximum value, the median is defined as the average of the curves maximizing depth.

\begin{minipage}[t]{0.95\textwidth}
\includegraphics[width=1.3\textwidth]{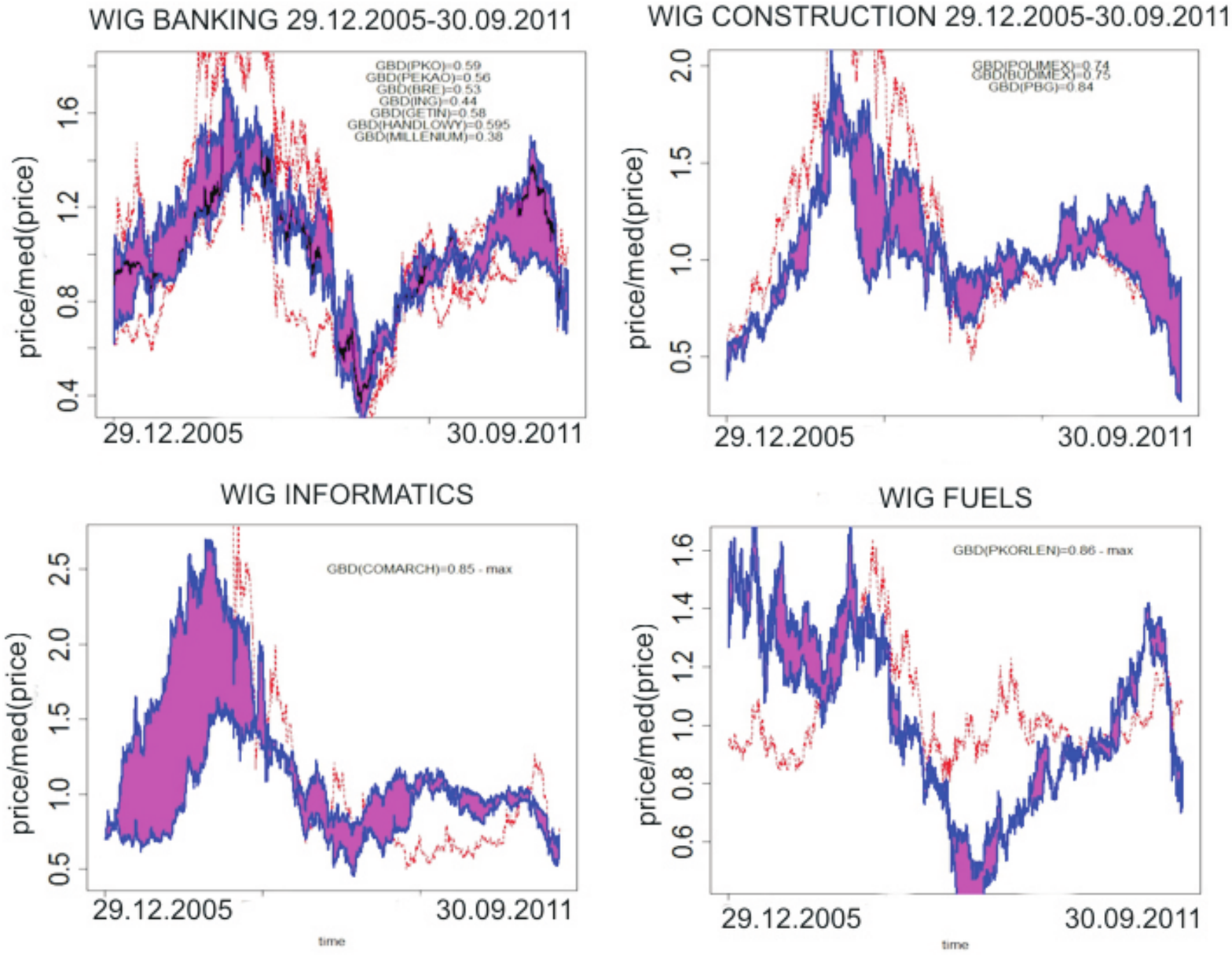}
\captionof{figure}{Functional boxplots (w.r.t. MBD) for Polish sector sub-indices in 2005 – 2011. Companies were considered with respect to a relative price of the stocks. Calculations made with fda R package (Ramsay et al., 2009 \cite{Ra})}
\label{fig:OG_1}
\end{minipage}

\begin{minipage}[t]{0.95\textwidth}
\includegraphics[width=1.3\textwidth]{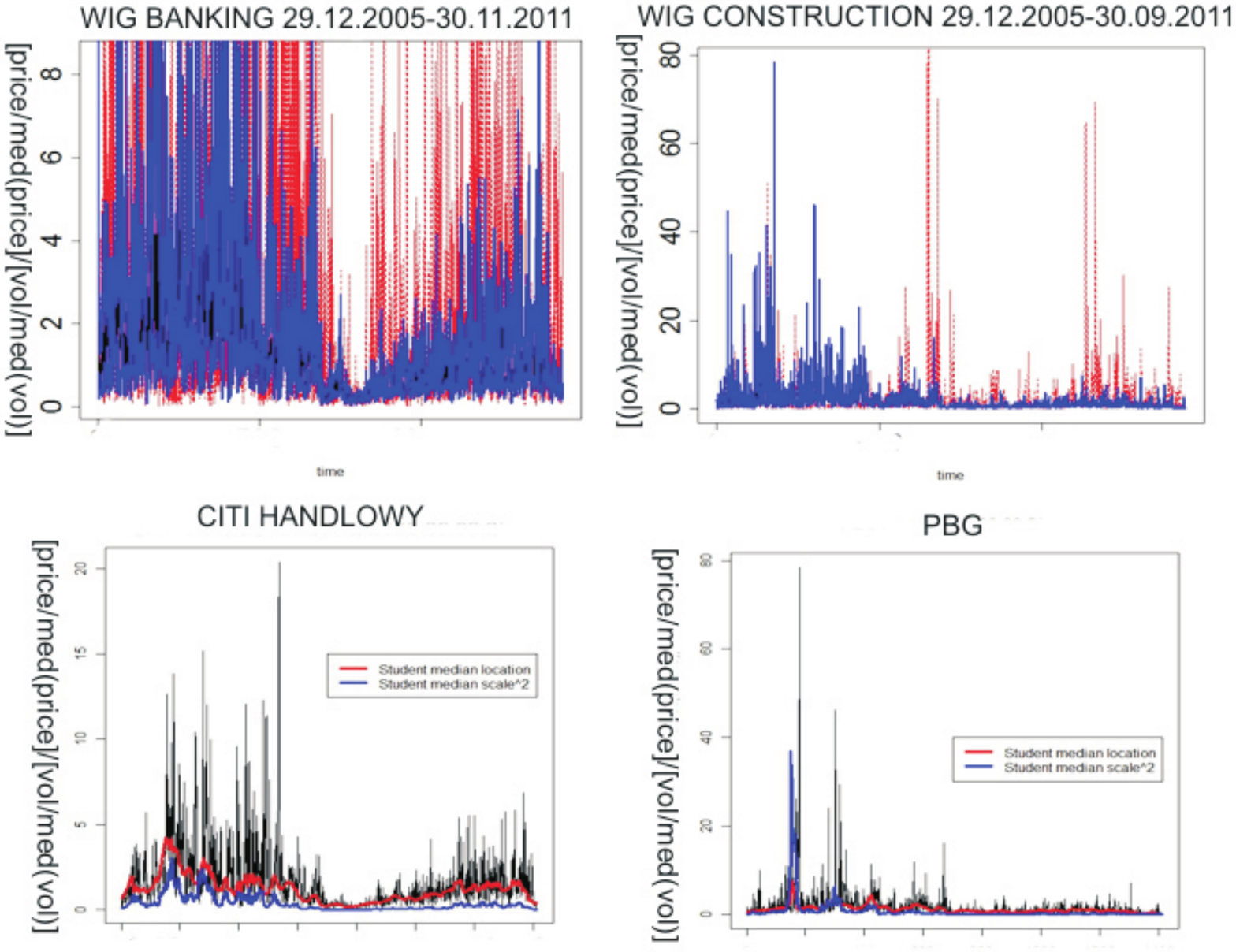}
\captionof{figure}{Functional boxplots for companies of two Polish sector sub-indices showing “a balance between supply and demand” (up) and companies, which are the median trajectories w.r.t. MBD (down)}
\label{fig:OG_2}
\end{minipage}
\vskip5mm

Figure 1 presents functional boxplots (with respect to the MBD) for Polish sector sub-indices in 2005 – 2011. Companies were considered with respect to a relative price of the stocks.
Figure 2 presents functional boxplots for companies of two Polish sector sub-indices showing “a balance between supply and demand” and companies, namely City Handlowy and PBG, which are the median trajectories of their sub-index with respect to the MBD. Both figures show the "energy" or dispersion of the considered quantity.
\\ Another example of statistical depth function, which we use further is a projection depth function -- we recommend this depth in a context of a balance between effectiveness, robustness and computational complexity. We define the projection depth of a point $x\in\mathbb{R}^d$ with respect to a sample $\mathbb{X}^n=\lbrace X_1,...,X_n\rbrace$ as
\begin{equation}
PD(x, \mathbb{X}^n)=\left[
1+\sup_{|u|=1}
|u^Tx-m(u^T\mathbb{X}^n)|/\sigma(u^T\mathbb{X}^n)
\right]^{-1},
\end{equation}
where $m$ and $\sigma$ are one-dimensional measures of the location and dispersion (for example a median (Med) or a median absolute deviation (MAD)), and $u^T\mathbb{X}^n=
\lbrace u^TX_n,..., u^TX_n \rbrace.$
Theoretical properties of this depth may be found in Zuo (2006) \cite{Zuo_proj}, for its R implementation see Kosiorowski and Zawadzki (2017) \cite{DepthProc}, and Liu et al. (2012) \cite{Zuo_et_comp_proj}.
We can define depth for vectors, matrices, functions, families of sets, geometrical objects (see Zuo and Serfling, 2000 \cite{Zuo_Serfling}). Depth functions yield nested contours of equal outlyingness regions and they are natural tools for peeling of data from outliers. 
\vskip1mm
PROPOSAL 1 (Robust estimator of a stress functional): Let $X^n=\lbrace x_1,...,x_n\rbrace $ denote a sample of $k\times 2$  configuration matrices (which may contain outliers). The matrices represent $k$ -- companies belonging to a sector stock index considered with respect to two variables: relative price and relative volume. In order to obtain a robust estimate of the average shape of the index (stress functional) and the variance of shape (a value of a strain force), we transform the matrices using vectorize operator, which transforms a matrix into a column vectors $vec(x_1),...,vec(x_n).$
Subsequently, we calculate sample projection depth $D(z,Z^n)$ for
$Z^n=\{ z_1=vec(x_n),...,z_n=vec(x_n)\}$
and throw away observations with depth $D(z,Z^n)$  smaller than a certain predefined threshold, e.g., $\alpha=0.1$. For the observations left, the ordinary Procrustes average and $SVAR$ measure are calculated.
\vskip1mm
PROPOSAL 2 (Robust analysis of a capital flow): Let $X_t^1=[x^1_1,...,x^1_{r_1}]$,..., $X_t^k=[x^k_1,...,x^k_{r_k}]$ denote $k$ -- companies belonging to $k$- sector sub-indices consisted of, correspondingly, $r_1,...,r_k$ companies considered with respect to the price and the volume and observed in time points $t=1,...,n.$ In order to obtain a robust estimate of the average shape (stress functional) and variance of the shape (a value of strain force), we perform the following steps:
\\ 1. For each of the $k$ sub-indices, i.e., for $r_k$ series of 
$$\{ price^i(t)/med^i(price^i(t))\}/\{ volume^i(t)/med^i(volume^i(t))\}\},$$ for $i=1,...,r_k,$  $t=1,...,n,$ we calculate López-Pintado and Romo (\cite{LopezRomo}) modified band depth (MBD) .
\\ 2. For each of the $k$ sub-indices, we choose a company being MBD median, i.e., its trajectory is the most representative, the most central w.r.t. MBD, for the studied period.
\\ 3. For these $k$- chosen MBD medians – time series representing companies chosen from $k$- sector sub-indices, we calculate ordinary Procrustes averages and $SVAR$ measures of the shape variability. \\
\vskip1mm
It is crucial, that due to statistical properties of used depth functions, i.e. affine invariance, both proposals lead to objects being shapes. 
\vskip1mm
In order to check a performance of the proposals, we conducted several simulation studies for configuration matrices generated from matrix analogues of multivariate normal, multivariate Student, and uniform distributions. We considered relatively small samples (50-150 observations of $8\times 2$ matrices) with and without up to 5\% additive outliers. Results of the simulations indicated consistency, reasonable finite sample robustness.
\section{Results of the empirical studies}
In Kosiorowski (2006) \cite{KosiorowskiPhD} five Polish sector stock sub-indices WIG-banking, WIG-construction, WIG-media, WIG-food and WIG-IT from 2005 were considered. The period was divided into two consecutive sub-periods (two half-years) and analyzed by means of the Proposal 1. The stocks belonging to the sub-indices were considered with respect do a daily change of the price and the volume. An equalities of two average shapes by means of robust version of Hotelling $T^2$ test were verified (for details of that test see Kosiorowski (2006) \cite{KosiorowskiPhD}). The study led to the following conclusions. 1. One observed capital flows related to the stresses in case of the WIG-media index. One, among other, observed a significant increment in shape variability (the stain) followed by increase in the overall volume (flow of the capital), significant deformation of the average shapes (local changes) and inequality of the average shapes (global changes). 2. For the index WIG-food and WIG-banking one observed capital flows caused by an activity of external forces (equality of the average shapes and equality of dispersions of shapes were simultaneous with the activity of the environment). 3. WIG-construction remained in a “dynamic equilibrium state”.
\vskip1mm
In 2011 we significantly changed and improved the approach. We realized that the financial crisis of 2007 year give us a unique chance for a verification of our concept. This time we focused our attention on the ratio $price/median(price)$ divided by the $volume/median(volume)$ for a certain representative stock. We interpreted this ratio in terms of “a dynamic equilibrium between supply and demand”, as a relation representing interplay between inner and external forces in the capital. Changes of this ratio can be interpreted as a perturbation in the space of economic values implied by or implying a capital flow. We considered eight sector stock indices: WIG-banking, WIG-construction, WIG-media, WIG-food, WIG-oil\&gas, WIG-telecom, WIG-IT and WIG-chemicals in the period 29.12.2005 – 30.09.2011. We divided the considered period into seven approximately equal sub-periods following one another. We used a MBD (L\'opez-Pintado and Romo, 2009 \cite{LopezRomo}) in order to choose from the index a representative particle of the capital, i.e., the landmark company was the median trajectory of the index. For these median trajectories of the indices – tools of the statistical theory of shape were applied. We calculated average Procrustes shapes, measures of dispersion of shapes $SVAR$ and likewise we calculated PTPS deformation of the average shapes in the consecutive sub-periods. Figure 3 presents estimates of the stress measure - average shapes - for six approximately equal consecutive sub-periods of the 2005 – 2011. Figure 4 shows PTPS of these average shapes for the consecutive sub-periods.\\
\begin{figure}
        \centering
        \begin{subfigure}[b]{0.44\textwidth}
                \centering
                \includegraphics[width=\textwidth]{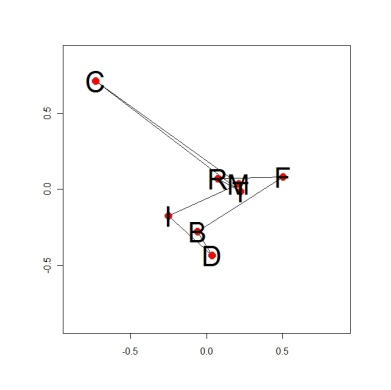}
                \caption{A1}
                \label{fig:A1}
        \end{subfigure}%
        \begin{subfigure}[b]{0.44\textwidth}
                \centering
                \includegraphics[width=\textwidth]{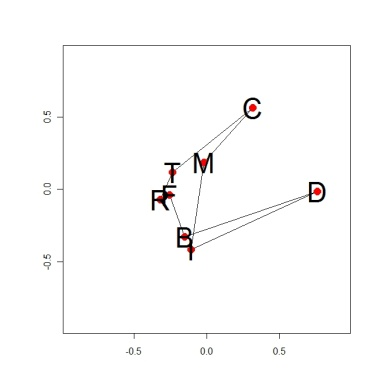}
                \caption{A2}
                \label{fig:A2}
        \end{subfigure}%

        ~ 
        \begin{subfigure}[b]{0.44\textwidth}
                \centering
                \includegraphics[width=\textwidth]{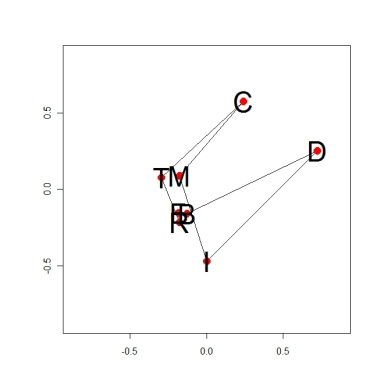}
                \caption{A3}
                \label{fig:A3}
        \end{subfigure}
        ~ 
        \begin{subfigure}[b]{0.44\textwidth}
                \centering
                \includegraphics[width=\textwidth]{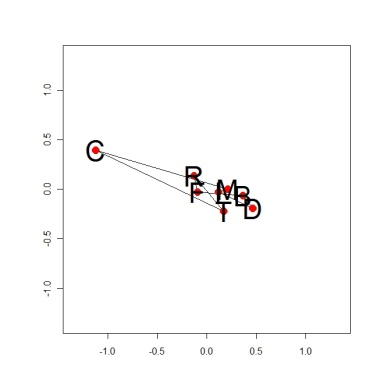}
                \caption{A4}
                \label{fig:A4}
        \end{subfigure}
         \begin{subfigure}[b]{0.44\textwidth}
                \centering
                \includegraphics[width=\textwidth]{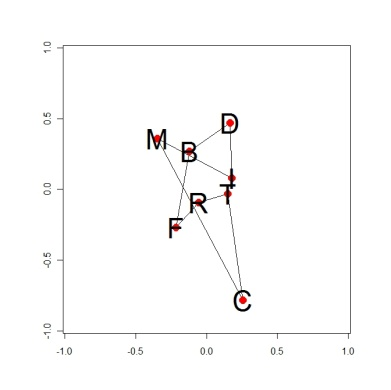}
                \caption{A5}
                \label{fig:A5}
        \end{subfigure}
         \begin{subfigure}[b]{0.44\textwidth}
                \centering
                \includegraphics[width=\textwidth]{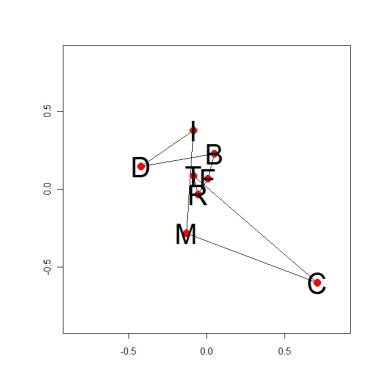}
                \caption{A6}
                \label{fig:A6}
        \end{subfigure}
\caption{Estimates of the stress measure - average shapes - for six approximately equal consecutive periods of the 2005 – 2011}
\label{fig:Wig_Shapes}
\end{figure}
\begin{figure}
        \centering
        \begin{subfigure}[b]{0.45\textwidth}
                \centering
                \includegraphics[width=\textwidth]{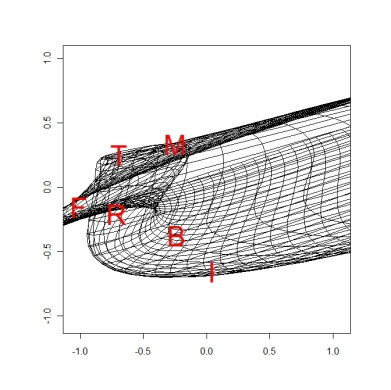}
                \caption{T1}
                \label{fig:T1}
        \end{subfigure}%
        \begin{subfigure}[b]{0.45\textwidth}
                \centering
                \includegraphics[width=\textwidth]{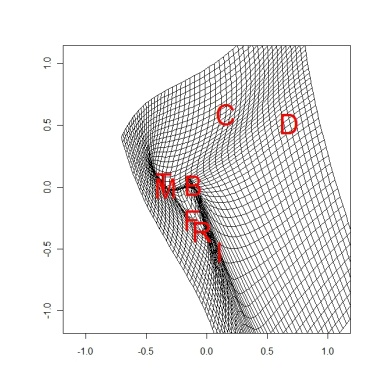}
                \caption{T2}
                \label{fig:T2}
        \end{subfigure}%

        ~ 
        \begin{subfigure}[b]{0.45\textwidth}
                \centering
                \includegraphics[width=\textwidth]{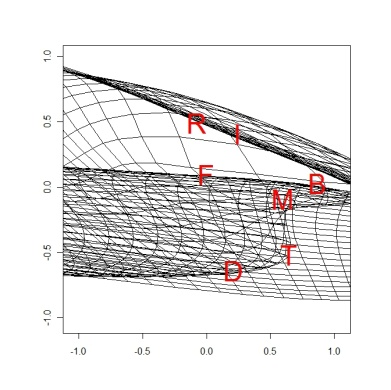}
                \caption{T3}
                \label{fig:T3}
        \end{subfigure}
        ~ 
        \begin{subfigure}[b]{0.45\textwidth}
                \centering
                \includegraphics[width=\textwidth]{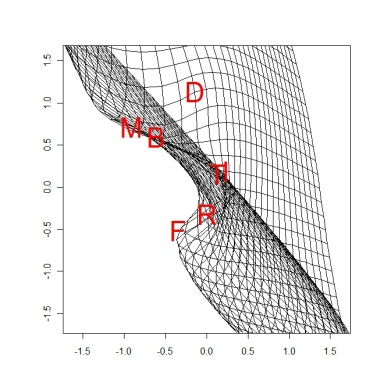}
                \caption{T4}
                \label{fig:T4}
        \end{subfigure}
         \begin{subfigure}[b]{0.45\textwidth}
                \centering
                \includegraphics[width=\textwidth]{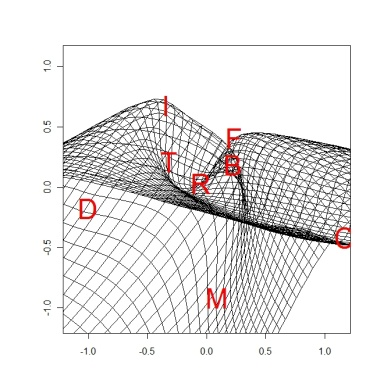}
                \caption{T5}
                \label{fig:T5}
        \end{subfigure}
         \begin{subfigure}[b]{0.45\textwidth}
                \centering
                \includegraphics[width=\textwidth]{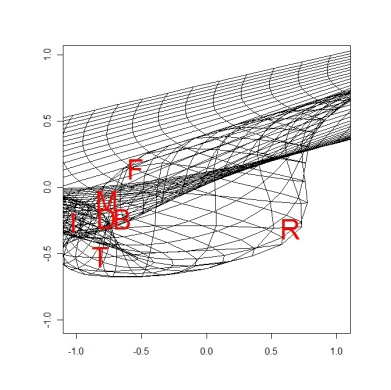}
                \caption{T6}
                \label{fig:T6}
        \end{subfigure}
        \caption{PTPS of consecutive stress functionals}\label{fig:PTPS}
\end{figure}

The financial crisis of the 2007 year manifested on the Figure 3(c) and 3(d) (values of the stress measure) and Figure 4(c) and 4(d) (PTPS deformation of representative particles of the capital stored on the stock market). Figures 5 – 16 present four WIG sub-indices with respect to proposed measures of a) direction of inflow/outflow, b) intensity of inflow/outflow, c) perturbations in the space of economic values. A comparison of Figures 3(c), 3(d) and 4(c) with Figures 5 – 16 lead us to the following conclusions: 1. We observed elastic deformation of the stock market. 2. We observed outflow of the capital from stock market which was related to the stresses. Stresses manifested in non-homogeneous behaviors of the sub-indices (WIG-construction and WIG-IT vs. WIG-banking and WIG-oil\&gas). 3. The energy of the market was absorbed by the stresses in the capital to the minimal value during the apogee of the crisis and then the energy was given back to the market in effect of the elastic recovery.

\begin{minipage}[t]{0.49\textwidth}
\includegraphics[width=0.99\textwidth]{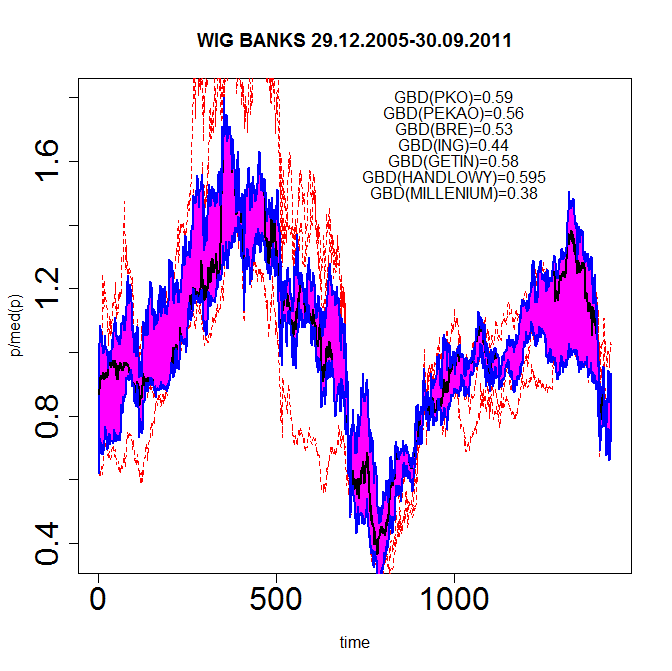}
\captionof{figure}{WIG-banking - relative price – inflow/outflow.}
\label{fig:R1}
\end{minipage}
\begin{minipage}[t]{0.49\textwidth}
\includegraphics[width=0.99\textwidth]{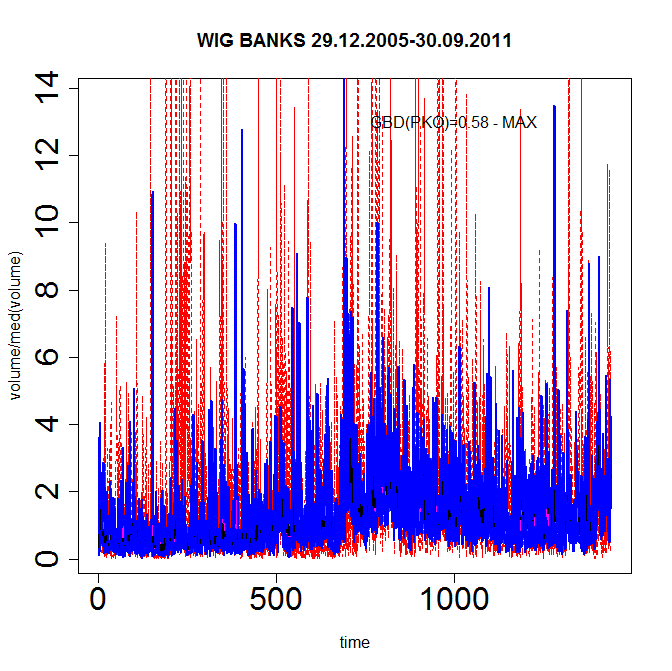}
\captionof{figure}{WIG-banking – an intensity of the flows.}
\label{fig:T2}
\end{minipage}

\begin{minipage}[t]{0.49\textwidth}
\includegraphics[width=0.99\textwidth]{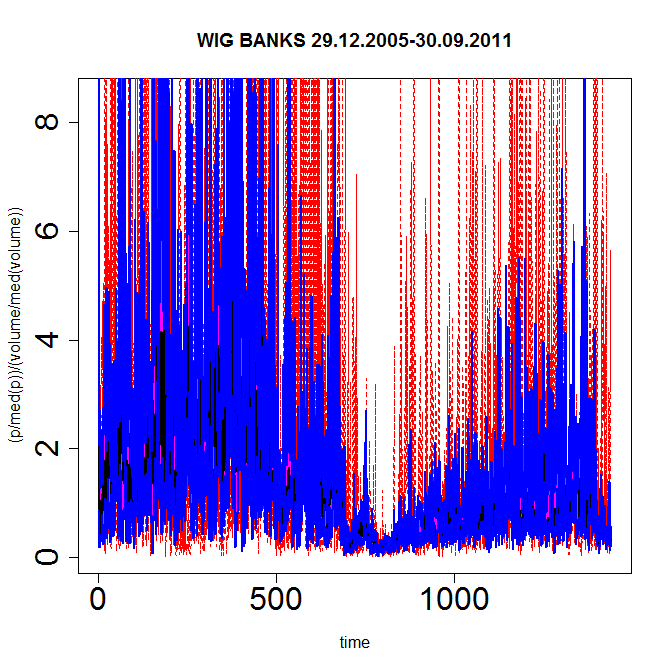}
\captionof{figure}{WIG-banking – perturbations in the space of economic values.}
\label{fig:R3}
\end{minipage}
\begin{minipage}[t]{0.49\textwidth}
\includegraphics[width=0.99\textwidth]{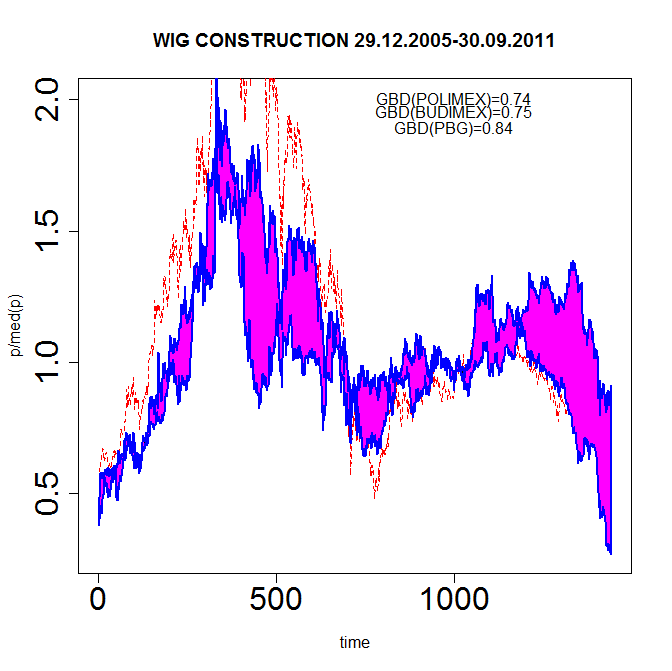}
\captionof{figure}{WIG-construction relative price – inflow/outflow. }
\label{fig:R4}
\end{minipage}

\begin{minipage}[t]{0.49\textwidth}
\includegraphics[width=0.99\textwidth]{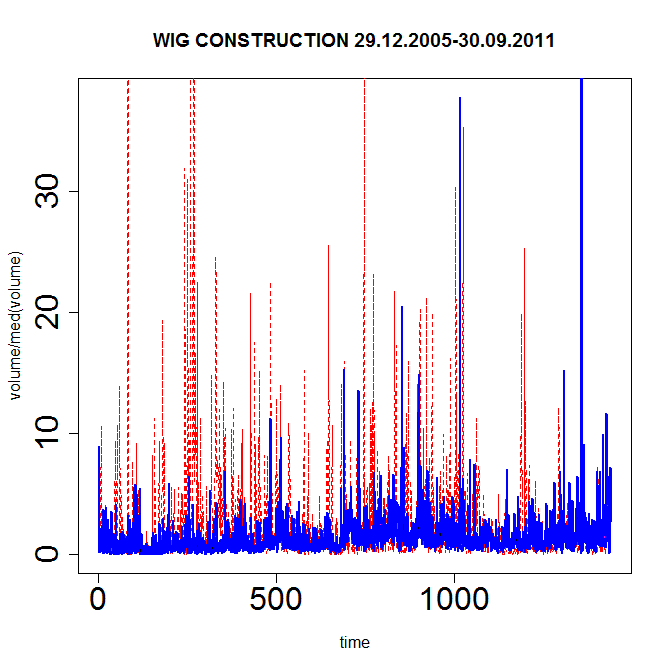}
\captionof{figure}{WIG-construction an intensity of the flows.}
\label{fig:R5}
\end{minipage}
\begin{minipage}[t]{0.49\textwidth}
\includegraphics[width=0.99\textwidth]{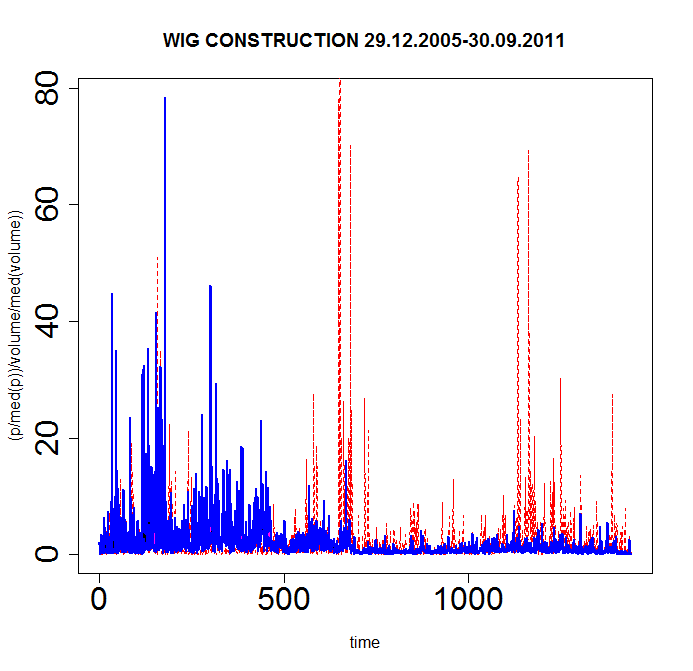}
\captionof{figure}{WIG-construction – perturbations in the space of economic values.}
\label{fig:R6}
\end{minipage}

\begin{minipage}[t]{0.49\textwidth}
\includegraphics[width=0.99\textwidth]{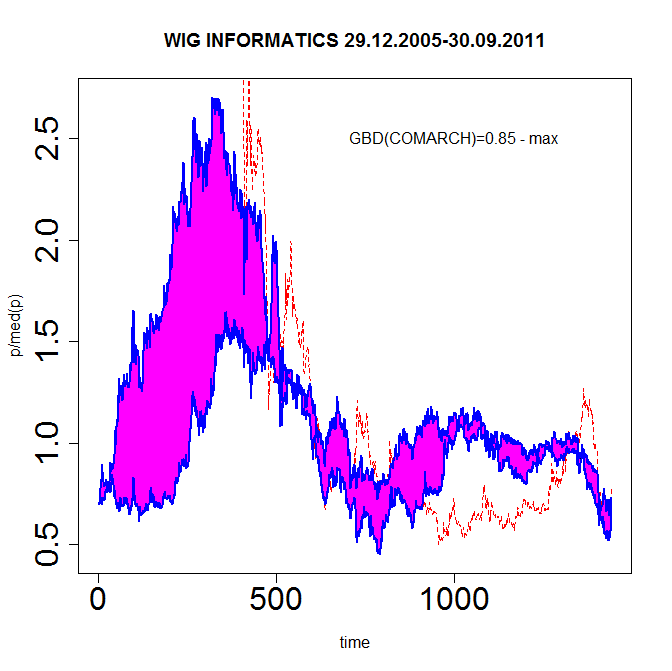}
\captionof{figure}{WIG-IT relative price – inflow/outflow.}
\label{fig:R7}
\end{minipage}
\begin{minipage}[t]{0.49\textwidth}
\includegraphics[width=0.99\textwidth]{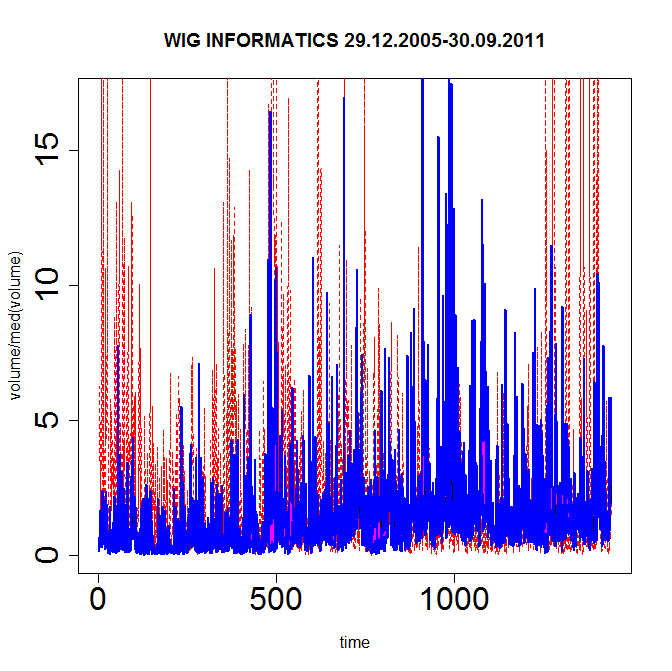}
\captionof{figure}{WIG-IT an intensity of the flows.}
\label{fig:R8}
\end{minipage}

\begin{minipage}[t]{0.49\textwidth}
\includegraphics[width=0.99\textwidth]{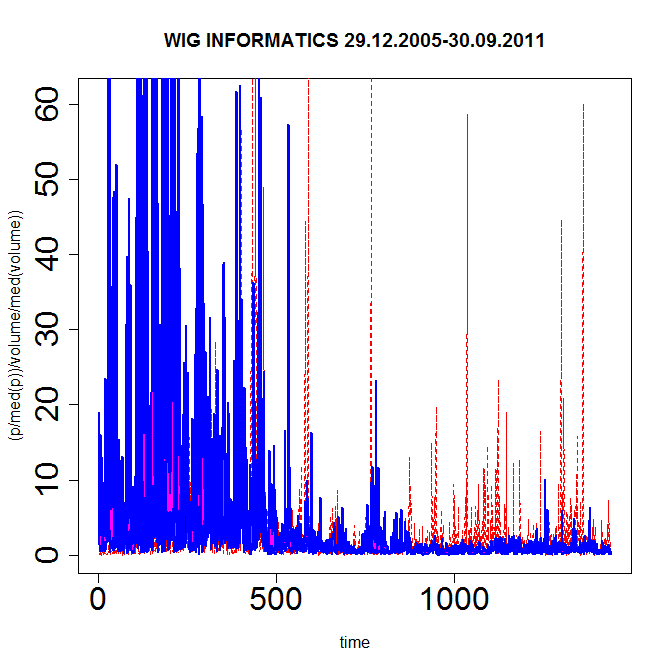}
\captionof{figure}{WIG-IT – perturbations in the space of economic values.}
\label{fig:R9}
\end{minipage}
\begin{minipage}[t]{0.49\textwidth}
\includegraphics[width=0.99\textwidth]{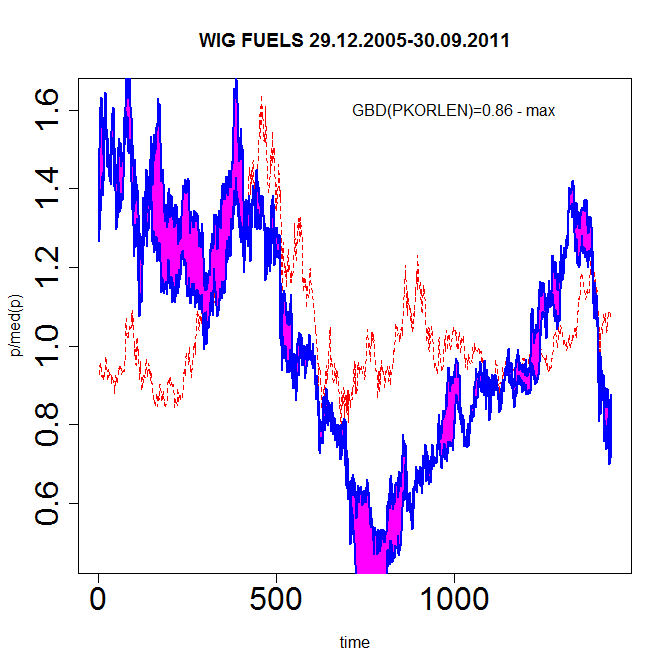}
\captionof{figure}{WIG-oil\&gas relative price – inflow/outflow.}
\label{fig:R10}
\end{minipage}

\begin{minipage}[t]{0.49\textwidth}
\includegraphics[width=0.99\textwidth]{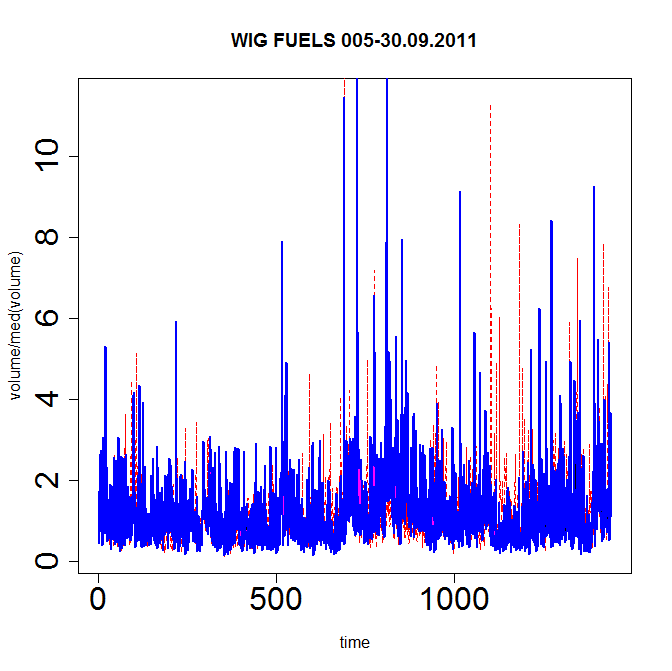}
\captionof{figure}{WIG-oil\&gas an intensity of the flows.}
\label{fig:R11}
\end{minipage}
\begin{minipage}[t]{0.49\textwidth}
\includegraphics[width=0.99\textwidth]{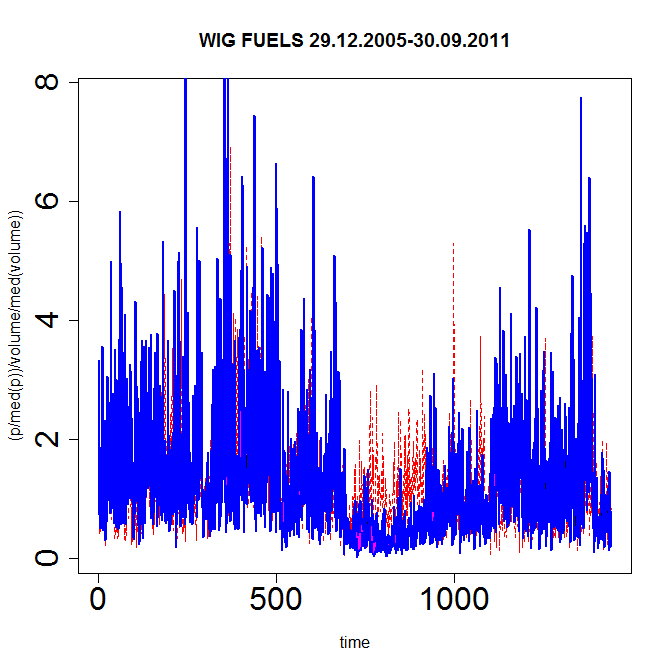}
\captionof{figure}{WIG-oil\&gas – perturbations in the space of economic values.}
\label{fig:R12}
\end{minipage}
\begin{figure}
        \centering
        \begin{subfigure}[b]{0.44\textwidth}
                \centering
            \includegraphics[width=\textwidth]{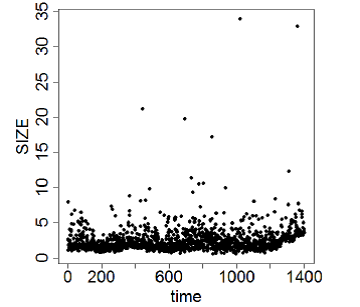}
                \caption{Centroid sizes of configuration matrices for eight representative companies belonging to Polish sector sub-indices in 2005 – 2011 considered w.r.t. the relative price and the relative volume.}
                \label{fig:V1}
        \end{subfigure}%
        ~
        \begin{subfigure}[b]{0.44\textwidth}
                \centering
                \includegraphics[width=\textwidth]{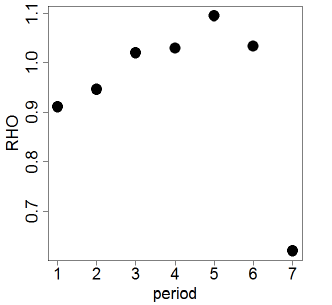}
\caption{Measures of shape dispersion for eight representative companies belonging to Polish sector sub-indices in seven consecutive sub-periods of 2005–2011.}
      \label{fig:V2}
        \end{subfigure}%
               \caption{"A temperature of a capital" (a) and sizes of a force inducing a capital flow between economic systems (b)}\label{fig:variability}
\end{figure}

Results of the empirical studies showed an increase of shape variability measure and hence the strain force in a period preceding the financial crisis ans sudden decrease after it (see Figure 17(b)). A time evolution of the stress functional presented on Fig. 4 show huge amounts of local stress in the capital in the considered period with "qualitatively" the highest value in apogee of the financial crisis. Figures 7 and 16 clearly show a gap in perturbations in a space of economic values in the apogee of the crisis - sudden "crash of a capital"    

\section{Conclusions}
The robust statistical framework for a measurement of stresses in the capital stored on the stock market was proposed. By means of the proposed tools, several empirical findings were presented. 
The most important finding is that stresses on a financial market expressed in terms of a deformation of the stress functional, and an increase of the shape variability measure are both increasing when an apogee of the crisis is approaching, and perturbations in the space of economic values are decreasing at the same time. 
In our opinion, the presented approach is worth further studying involving simplification of the statistical part and formalization of the conceptual issues concerning the notion of the capital. The approach may be useful in predicting financial crashes in a future. 
\vskip1mm
\textbf{Acknowledgements}
JPR and DM's research has been partially supported by the AGH UST local grant no. 11.11.420.004
and DK's research by the grant awarded to the Faculty of Management of CUE for preserving scientific resources for 2017 and 2018 year.

\end{document}